\documentclass[runningheads]{llncs}
\usepackage{wasysym}
\usepackage{graphicx}
\usepackage{cite}
\usepackage{multirow}
\begin{document}
\title{A Conceptual Framework for Implicit Evaluation of Conversational Search Interfaces }
\titlerunning{Implicit Evaluation of Conversational Search Interfaces}
\author{Abhishek Kaushik%\inst{1}%\orcidID{0000-1111-2222-3333} \and
\and 
Gareth J. F. Jones} %\inst{2}}%\orcidID{1111-2222-3333-4444}}
% \and Third Author\inst{3}\orcidID{2222--3333-4444-5555}}
%
\authorrunning{A. Kaushik et al.}
%\titlerunning{Workshop on Mixed-Initiative ConveRsatiOnal Systems 2021}
%\fancyfoot{Workshop on Mixed-Initiative ConveRsatiOnal Systems 2021 }
%\footnote{Workshop on Mixed-Initiative ConveRsatiOnal Systems 2021 }
%\fancyfoot[LE,RO]{Workshop on Mixed-Initiative ConveRsatiOnal Systems 2021 }

\institute{ADAPT Centre, School of Computing Dublin City University, Ireland \\
\email{abhishek.kaushik2@mail.dcu.ie, Gareth.Jones@dcu.ie}}
%\institute{Adapt Centre, Dublin City University, Ireland} %\and
%Springer Heidelberg, Tiergartenstr. 17, 69121 Heidelberg, Germany
%\email{lncs@springer.com}}\\
%\url{http://www.springer.com/gp/computer-science/lncs}} % \and
\maketitle              % typeset the header of the contribution
\begin{abstract}
Conversational search (CS) has recently become a significant focus of the information retrieval (IR) research community. Multiple studies have been conducted which explore the concept of conversational search. Understanding and advancing research in CS requires careful and detailed evaluation. Existing CS studies have been limited to evaluation based on simple user feedback on task completion. We propose a CS evaluation framework which includes multiple dimensions: search experience, knowledge gain, software usability, cognitive load and user experience, based on studies of conversational systems and IR. We introduce these evaluation criteria and propose their use in a framework for the evaluation of CS systems. 
%This paper discussed, how conversational search agent would be evaluated to provide a framework that would support CS system in future.%CS systems are the prospective intelligent system that supports the users in the search process and reduce the cognitive load which can be initiated by the multi modal signals (speech and text).  % This study would support in defining the concept of conversational information retrieval precisely and completely. Based on our analysis, it is concluded that our framework would be the potential evaluation metric in Conversational Information Retrieval (CIR). 

\end{abstract}

\section{Introduction}

Recent progress in artificial intelligence has brought tremendous advances in conversational systems and information retrieval (IR). This has led to increasing interest in Conversational Search (CS) using conversational engagement to complete IR tasks \cite{radlinski2017theoretical}. CS presents opportunities to support users in their search activities to improve the effectiveness and efficiency of information seeking, while reducing their cognitive load. A number of studies have been conducted to examine the concept of the CS \cite{radlinski2017theoretical,Abhi}. 
%To define the concept of the conversational search, it is useful to explore and investigate the opportunity of evaluation metric of CS in the current context.
We believe that greater insight into the processes and potential of CS can be achieved using a detailed evaluation, and that this can help in advancing
%defining 
and understanding the exploitation 
%implementation 
of the paradigm 
%concepts 
of CS. 
%This 
These insights will help in enhancing proposed models and theories of CS. 
%We believe, the evaluation techniques framework will open much insight to design feasible Conversational search agent. 

This paper overviews \footnote{Accepted in ECIR 2021 Workshop on Mixed-Initiative ConveRsatiOnal Systems \url{https://micros2021.github.io/}}  the current methods and techniques used in
%by for 
the evaluation of
%in 
conversational systems in different dimensions, and use them  
%The paper will also include a prospective approach for evaluation of conversational information retrieval.
%The evaluations measures of CS contain two types of methodology such as quantitative and qualitative. The main focus of this paper 
to define a framework for the definition and utilization of evaluation metrics for CS.

%In total there should be 5 pages
%\textbf{Introduction (1/2 page)}

\section{Background}

%\subsection{Evaluation on Conversational search}

This section introduces existing work examining the evaluation of interactive IR, conversational systems and CS.

\subsection{Evaluation of Interactive Information Retrieval}

Interactive IR
%Information Retrieval 
(IIR) studies user interaction with search systems. The evaluation methods for IIR can be 
%are 
broadly classified into four major classes: contextual, interaction, performance and usability \cite{kelly2009methods}.

\begin{itemize}

\item \textbf{Contextual}: This measures 
%explain 
the context in which a search and interaction activity occurs, and characterizes the subject and their information need. Characteristics of subjects include: age, sex, search experience, etc. Characteristics of information need focus on information seeking situations such as the subject's background knowledge, subject familiarity with the search topic, etc. These measures basically describe the context in which the information search occurs \cite{ingwersen2005jarvelin,dourish2004we,kelly2009methods}.
%Ingwersen \textit{et al.} \cite{ingwersen2005jarvelin} did a comprehensive study of the context in information seeking and retrieval . Dourish \cite{dourish2004we} explained the theoretical examination of the concept. 
%The contextual measure often used in association with other types of measures.

\item \textbf{Interaction}: This focuses on characterizing the interactions between a
%the 
search system and the user.
%subject. 
This also includes the interactive search behaviour of the user, such as the length of each query, the number of queries entered, and the number of returned documents read, etc\cite{kelly2009methods}. %This is one of the most important components with respect conversational information retrieval 

\item \textbf{Performance}: This focuses on results obtained from the user's interaction with a search
%the 
system, such as calculating %presenting 
%the 
%relevant documents, 
precision, mean average precision, and recall of retrieved documents. According to Saracevic \cite{saracevic1975relevance}, these performance measures depend on the concept of relevance, the user's criteria of relevance assessment, and the techniques used for measuring the relevance.  
%The third set of measures are performance-based measures related
%to the outcome of the interaction, such as the number of relevant documents saved, mean average precision, discounted cumulative gain, and discounted cumulative gain. It also include time based measure, Informativeness and cost based measure. The criteria to select the performance measure depend upon the expectation of the IIR task.  
%The performance measures are explained in details in table 1 \cite{kelly2009methods}.

%The performance measure depends upon the method which is used to measure the relevance of the search result.
%ccording to the study of Kelly \cite{kelly2009methods}, the performance measure investigated and depended upon the following areas as shown in figure \ref{performance} \cite{saracevic1975relevance}.

\item \textbf{User-feedback}: This  captures the user's feelings and experiences of their interactions with the search system. This measure is also referred to as 
%a 
``usability'' and is divided into multiple dimensions \cite{iso19989241}. According to the  International Organization for Standards  (ISO), the key dimensions of usability are: effectiveness, efficiency, and satisfaction \cite{iso19989241}. %as shown in Table \ref{KeyDimesnion}.

%\begin{enumerate}
 %   \item Effectiveness :
%It is the \textit{accuracy and completeness with which subject achieve the specified task.} In general, a system is effective if it helps subject to fulfill their search tasks. 
%\item Efficiency:
%It is in the context of resources spend during the search task on effective search system. In general, a system is efficient if it enables subjects to fulfill their tasks with minimum expense and effort. 
%\item Satisfaction
%It is the \textit{freedom from discomfort, and positive attitudes of the user to the product [141]}.  Satisfaction is the feeling of contentment that subject experience after finishing the search task.
%\end{enumerate}

%The choice of any measure and its interpretation depend upon the nature of the task and the subject expectations from the search system.  %for an instance, the subject need to explore the topic then the major focus would be on the number of interactions and in contrast if the subject is looking for the specific answers then finding one or small number of highly relevant documents would be the key and system will be measured on based of high precision result to answer the specific question.  

\end{itemize}

\subsection{Evaluation of Conversational Systems}

Conversational engagement is currently being investigated as the mode of engagement 
%interface 
for many human-machine applications. Research in conversational systems has proposed 6 dimensions for the evaluation of conversational agents \cite{radziwill2017evaluating,ChatbotMetric}.

\begin{itemize}

\item \textbf{Extensive Capabilities}: Conversational systems should provide 
%a good texting 
an error-free environment to the user \cite{radziwill2017evaluating,morrissey2013realness,ChatbotMetric}. This should include spell checking and auto-correction to support error free expression by the user of statements and questions. In addition, conversational systems should use an appropriate combination of multimedia and text content \cite{eeuwen2017mobile,wilson2017jobs}. %This is a quantitative variable.

\item \textbf{User Interaction and Engagement}:% is also a qualitative measure variable that deals with user interaction. This 
This measure contains the following parameters:
%such as 
%capable 
capability of initiating
%initiative the 
conversations, maintaining 
%the 
conversational engagement, identifying and distinguish target users, etc.\cite{Anu}. %, respond meaningful response, support user in navigation of the conversational system and designed to answer the frequently answered questions (FAQ) based on the history and personal information of the user \cite{morrissey2013realness, ramos2017screw}. The above-mentioned parameters would be a good measure of the retention features of the conversational system \cite{radziwill2017evaluating, ChatbotMetric}. 

\item \textbf{Response speed}: The response speed of conversational agents should be sufficiently fast to prevent user frustration \cite{ChatbotMetric, radziwill2017evaluating, thieltges2016devil}. 
 
\item \textbf{Functionality}: 
%This measures 
The overall capabilities of the system as measured qualitatively by users based on multiple variables such as the richness of media supported, navigation tools provided to support users, multi-modality of engagement, etc \cite{Chatbotgood, morrissey2013realness}. 

\item \textbf{Interoperability}: This defines the ability of a conversational system to exchange and make use of information.  A standard conversational system enables the user to engage with multiple media channels \cite{ChatbotMetric}. 

\item \textbf{Scalability}: This is a  quantitative measure of the scope of the system to support multiple users. For example, the number of users supported by the conversational system at the same time, types of 
%the 
server that can accommodate the conversational system, database size, etc \cite{ChatbotMetric}. 

\end{itemize}

%Additionally, 
A well known framework for evaluation of conversational systems is the Paradise framework
%. This was developed 
\cite{walker1997paradise}. This includes measure such as:
%including 
task success, conversation efficiency (task duration, 
%the 
dialogue turns), conversation quality (response accuracy), and user satisfaction (ease of the task, user behaviour). However, 
%but is limited to the essential components to 
Paradise is of limited value for the evaluation of of conversational search systems, since it does not include important factors such as cognitive load and the knowledge gain during the search process. 
%and the metric associated with Paradise does not consider
%is not considering 
%factors which can affect the search process.  
Moreover, Paradise 
%was 
focuses on a
%the 
goal-oriented agent, which is different from a
%the 
search-oriented task for which the requirements can change as the user progresses through the search process. Hence, Paradise 
%it would 
is not suitable for evaluation of 
%not be the perfect fit for the 
conversational search systems.

\subsection{Evaluation of Conversational Search}

%GJ: YOUR IDEA?
In recent years, a number of studies have been conducted on CS 
%conversational search 
systems and interfaces. There studies can be broken down into four different approaches: 
%a) 
using existing conversational agents \cite{landoni2019sonny,Avishek}, 
%b) 
using human experts \cite{trippas2017people, TRIPPAS2020102162},
%c), 
%perceived experiment 
human covertly taking the role of automated agent (Wizard-of-Oz approach), \cite{avula2018wizard,dahlback1993wizard, mctear2016evaluating}, and 
%d) 
using rule-based or machine learning conversational interfaces \cite{kaushik2020interface, DBLpKaushikLJ21}. 
%It has been observed in most of the studies, the 
Evaluation in most of this work
%of the system was 
is limited to NASA TASK Load \cite{hart1988development} or SUS (Usability) \cite{brooke1996sus} or both \cite{avula2019embedding, dubiel2018investigating}. Some of these studies have also investigated 
%the 
sentiment 
%analyse 
\cite{dubiel2018investigating} 
%approach on 
in the user response to examine
%evaluate 
the relationship between the user’s mood 
%to co-related with 
and task success. %Studies also purpose to evaluate audio summaries to indirectly evaluate its quality \cite{trippas2017crowdsourcing}. 

There is also active research exploring the use of evaluation benchmarks for CS. 
%Most of these studies are covering limited dimension of the search process. 
Most notable is the 
%In these studies such as  
TREC CAsT track \cite{aroradcu,dalton2019cast}. An alternative interpretation of CS is examined in the FIRE RCD track \cite{AbhiFire}.
%, there is no scope of query building with incremental learning based on user conversational experience. Another limitation is query building based on the combination of background knowledge of user and information gain during the conversation. The extraction of a query from dialogue or a series of questions without knowing the pre-existing knowledge of the user is not realistic and practical.  
These tasks have examined query interpretation and response in the context of conversational engagement and query extraction from conversations respectively. In both cases, evaluation is largely limited to traditional approaches used for IR tasks.

Evaluation of 
%these 
interfaces in CS is 
%actually 
a highly complex topic involving multiple dimensions including the user's background knowledge of the search topic, their familiarity with the conversational agent, etc. In this
%current 
paper, we propose
%ing the 
a framework for the evaluation of CS
%evaluate conversational search 
interfaces using
%into 
five 
%different 
dimensions: user search experience \cite{Abhi}, knowledge gain \cite{wilson2013comparison}, cognitive and physical load \cite{hart1988development}, usability of the interface software \cite{lewis1995ibm} and user experience \cite{hinderks2018benchmark}. 

\section{Framework for the Implicit Evaluation of Conversational Search Interfaces}

%This section discusses the vital factor responsible to evaluate the CS interface, design, development, implementation and analyzes of the evaluation framework.
%Based on the Dagstuhl Seminar  the researcher convenience with including several evaluation dimensions, such as: user, retrieval, and dialogue for conversational search. The report
%It 
%suggests that the dimensions might have an overlap with those of IIR.
%Interactive Information Retrieval.  

In this section we introduce our
%This 
framework for the evaluation of CS 
%conversational search with
which
%specifically 
focuses on 
%the 
multiple 
%user 
dimensions relating to the user.

Most 
%of 
%the 
%conversational search 
CS studies so far reported have
%on were 
focused on user search experience of the task or the usability of
%experience on 
CS systems.  This has provided feedback 
%was 
focused on user search experience. In our 
%proposed 
framework,
%current proposal,
%study, 
evaluation 
%metrics are
%has been drafted 
is based on 
%the 
five factors responsible for the needs of 
%the 
CS outlined in the next section.
%conversational.
%search including usability. 
An approach of this sort is also advocated in the summary report from the Dagstuhl Seminar on CS
%conversational search 
\cite{anand}.

\begin{table*}[t]
\vspace{-2ex}
\centering
\begin{tabular}{l|l}
\hline
      &                                               Topics   (0 (very low)- 7 (very high)) \\
      \hline
      & Background Knowledge                                \\
Search Formulation (Per-Search)     & Interest in Topic                                        \\
 & Anticipated Difficulty                                    \\
\hline 
      & Actual Difficulty                                         \\
 %     & Helpfulness Highlighted                                  \\
 Content Selection     & Text Presentation Quality \\
      & Average number of docs viewed per search \\ 
      & The usefulness of Search results                           \\
 & Text Relevance    \\ \hline 
      & Cognitively Engaged                                     \\
       & Suggestions Skills                                     \\
      
 Interaction with Content     & System Understanding Input                                \\
      & Average Level of Satisfaction                                   \\ \hline 
      & Search Success                                          \\
     Post Search & Presentation of the Search Results                        \\
       & Expansion of knowledge after the search\\
       & Understanding about the Topic \\
        \hline
      
\end{tabular}
\caption{Flowchart of characteristics of the search process \cite{vakkari2016searching} by 
%the 
change in knowledge structure.}
\label{FLowchart}
%\vspace{-2ex}
\end{table*}

\subsection{Essential Factors for Conversational Search}
\label{Essential Factors for Conversational}

We identify the following 
%are the 
essential factors for the 
%to 
evaluation of CS interfaces.
%conversational search interface. 
%\begin{figure}
 %   \centering
  %  \includegraphics[width=\textwidth]{figure/eval.png}
   % \caption{Five dimensions for evaluation}
    %\label{fig:eval}
%\end{figure}

\begin{enumerate}

\item Cognitive Load: Conventional search can impose a significant cognitive load on the searcher \cite{kaushik2019dialogue}. An important factor in the evaluation of conversational systems is measurement of the cognitive load experienced by users while using the system.

\item Cognitive Engagement: It has been observed that users get frustrated if they find it difficult to search about their 
%specific 
topic of interest to satisfy their information need. Frustration can reduce the user's engagement with a search system and their associated effort to locate relevant information.% And user ends the search with the superficial information gain about the topic. 

\item Search as Learning: Learning while searching is an integral part of the information seeking process.  %It is anticipated during the search process; the users learn about the topic at each phase.  In general, there are three phases which are equally important for search prospective. 
In our current study, we propose 
%prepared
a metric which breaks this down into three phases 
%with vertical alignment based on the concept of search as learning 
\cite{vakkari2016searching, Abhi}, as shown in Table \ref{FLowchart}. It has been observed in conventional search that %due to 
high cognitive load and lower cognitive engagement impacts on user learning during the search process \cite{gwizdka2010distribution}.
%effects.
%the process of learning during the search. 
%This scenario could be very expensive for the resource point of view and overall reduce the moral of the user to learn while search. 

\item Knowledge Gain: 
%One can observe the process of learning by studying the parameters which effect the learning while search. Learning not only focused on search topic but also the about the other subjects during the multiple passes of search session. 
%Its very important in the search process, that the learning transformed into factual knowledge about the search topic. 
Satisfaction of the user's information need is directly related to their knowledge gain about the search topic. Knowledge gain can be measured based on recall of new facts gained after the completion of the search process \cite{wilson2013comparison}. 

\item User Experience (UX): Another important aspect that needs to be considered for evaluation of CS
%a conversational search 
systems is UX. 
% This dimension focus on users experience and their expectation in terms of usability of any product.  
%Modern smart speakers like Alexa or Google assistant are interactive and easy to use system which raised the expectations of users concerning the quality of conversational system. It has been observed the user’s expectations are quite high on user interfaces in today’s era. 
User experience is generally classified into two aspects: pragmatic and hedonic, which can be further divided into six components:
%scales such as 
attractiveness, perspicuity, efficiency, dependability, stimulation and novelty \cite{hinderks2018benchmark}. These factors all provide
%the above 
measures of
%shows the 
user 
ease of use and the dependability of a conversational system  \cite{hinderks2018benchmark}.  

\item Software Usability: CS
%Conversational search 
studies generally do not explore the dimensions of 
%the 
software usability. However, it is important to understand the challenges and opportunities of conversational systems on the basis of software requirements analysis. This allows a system to be evaluated based on real life deployment and to identify
%outline the 
areas 
%need to be 
for improvement. Lower effectiveness and efficiency of a software system can increase cognitive load, reduce engagement and act as a barrier in the process of learning while searching. %Any software which score low usability score, is not consider efficient and effective for the users. 

\end{enumerate}

%In the context of factors effective CS, we have divided the evaluation metric into 2 dimensions i.e. . % as shown in the figure \ref{fig:eval}. 
%Careful consideration of all dimensions would lead us to refine and validate the CS evaluation metric. 

\begin{figure}
    \centering
    \includegraphics[width=\textwidth]{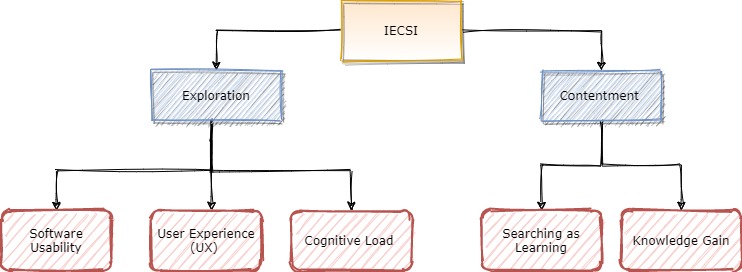}
    \caption{Implicit Evaluation of a 
    %for 
    Conversational Search Interface (IECSI).}
    \label{fig:IECSI}
\end{figure}

\subsection{Designing our Conceptual Framework}
\label{coneptual}

In our work on CS, we have conducted 
%examined 
multiple studies of CS and the use of
%conversational search and 
associated conversational agents. In conjunction with this, we have also conducted four user studies on conventional IR systems, a commercial conversational system (Alexa Echo Show) and conversational search interfaces for
%highly 
complex search tasks. In these studies, we have examined user behaviour and user
%its 
expectations with respect to CS.
%the conversational search. 
Based on our investigations \cite{Abhi, kaushik2019dialogue, aroradcu,AbhiFire, kaushik2020interface}, we propose
%have drafted the implicit 
an evaluation framework 
%metric 
for CS 
%conversational search 
interfaces. The framework
%metric 
is divided into two segments:
%including
%such as 
Exploration 
%segment 
and Contentment, 
%segment 
as shown in Figure \ref{fig:IECSI}. This section outlines the combination of
%include 
%the 
standard questionnaires of multiple dimensions 
%to combined together 
which
%to 
form 
%Implicit 
our proposed Implicit Evaluation for Conversational Search Interface (IECSI). The details are as follows: 

\subsubsection{Explore Segment:}

This segment focuses on exploring and experiencing CS %conversational search 
interfaces, and 
%This segment evaluates the extensive capabilities and outline the expectations of the users. 
%This segment 
is classified into three components: 
%a) 
Software Usability, 
%b) 
User Experience, and 
%c) 
Cognitive Load.   

\begin{enumerate}
    \item Software Usability:
%    Usability is an
%%is one of the 
%important evaluation metric for IECSI. 
Usability is an important consideration for the evaluation of interactive software.
%The  psychometric evaluation 
the IBM Computer Usability Satisfaction Questionnaires enables psychometric evaluation from the perspective of the user, and 
%is %considered for framework, 
is
%are
known as the Post-Study System Usability Questionnaire (PSSUQ) \cite{lewis1995ibm}. 
The PSSUQ includes four dimensions: overall satisfaction score (OVERALL), system usefulness (SYSUSE), information quality (INFOQUAL) and interface quality (INTERQUAL), which includes sixteen parameters.
%is 
%a Psychometric Evaluation for 
%the 
%software from the perspective of the user \cite{lewis1995ibm} known as the Post-Study System Usability Questionnaire (PSSUQ) Administration and Scoring. The PSSUQ was
%has been 
%evaluated using
%into the 
%four dimensions:
%such as: 
%the 
%overall satisfaction score (OVERALL), system usefulness (SYSUSE), information quality (INFOQUAL) and interface quality (INTERQUAL), which include sixteen parameters.% On each dimension, the conversational 
%chatbot 
%search interface outperformed the conventional search interface. The grading scale lines between 0(very low)- 7(very High). We 
%have 
%compared the mean difference of both 
%the 
%systems on all six parameters. In all 
%the 
%aspects, subjects experienced less task load when using the conversational
%in chatbot 
%interface. %, as shown in Table \ref{Post-Study System Usability Questionnaire}. 
%\begin{figure}
 %   \centering
  %  \includegraphics[scale=0.5]{figure/UEQ_S.PNG}
  %  \caption{User Experience Questionnaire \cite {hinderks2018benchmark}}
   % \label{fig:USE}
%    \vspace{-2ex}
%\end{figure}

\item User Experience:
%To ensure a conversational search system to provide reasonable User Experience (UX), it is critical to have a measurability which define the user insight about the system.  
%The other dimension included in 
%considered for 
%the IECSI framework is UX. 
UX
%This 
is measured using a questionnaire for interactive product known as the User Experience Questionnaire (UEQ-S) \cite{hinderks2018benchmark,laugwitz2008construction,schrepp2017design}. %This questionnaire contains 26 parameters. 
This questionnaire also enables us to analyse and interpret outcomes by comparing against 
%with 
a 
%huge 
benchmark dataset of outcomes for other interactive products. This questionnaire also provides us with the opportunity to compare 
%to 
interactive products with each other. 
%For specified purpose, a brief version (UEQ-S) has been prepared which have 8 parameters to be considered \cite {hinderks2018benchmark}.  UEQ-S has been preferred for the conversational search interface based on the practical scenarios. For an example, user fill the experience questionnaire after finishing the search task, if there too many items to ask, then user might not fill the answer seriously or refuse to answer (as the user has finished the search task and in the process of leaving or start the next task, so the motivation to invest more time on feedback would be difficult).
UEQ-S contains two meta quality dimensions: pragmatic and hedonic. Each dimension contains four
%4 
different parameters 
%each 
as shown in the Table \ref{tab:User Experience}. Pragmatic quality 
%would 
explores the 
%realistically 
usage experience of 
%the 
a conversational search system. Hedonic quality 
%would 
explores the pleasantness of using the system. 
%Subjects can rate each parameter on a 7-point Likert scale \cite{joshi2015likert}. %Moreover, the average value (mean) of the eight parameters is used as an overall UX value.  Each parameters of the Questionnaire consist of a pair of antonyms as shown in Figure \ref{fig:USE}. Subjects can rate each parameter on a 7-point Likert scale \cite{joshi2015likert} % where -3 (strong negative) and + 3 (strong positive). All the parameters with their respective dimensions are shown in Table \ref {tab:User Experience}. The design process of USQ-S can be found in the study \cite{schrepp2017design}.

\item Cognitive Load: %In terms of cognitive load, the user was
%has been 
%asked to evaluate the conventional 
%traditional and 
%interface and MCSI 
%system 
%in
%on 
%6 dimensions \cite{hart1988development}.
%One of the 
An important consideration in the evaluation of CS interfaces is their impact on the user's cogntive load 
%requirements for the conversational search interface is to reduce
%workload 
during the search process. 
%Workload is directly propositional to cognitive load.  
To measure the user's workload, the NASA Ames Research Centre %did the investigation for 3 years including 40 different laboratories and 
proposed the NASA Task Load Index \cite{hart1988development}. This is a multi-dimensional rating procedure which provides a measurement of the overall workload during a
%the 
process or event. This workload is classified into six subscales: mental, physical, temporal, own performance, effort and frustration. Out of these six dimensions, three 
%dimensions 
are related to the demand imposed on the subject due to the task (mental, physical and temporal) and the remaining three to the interaction of the subject with the system (effort, frustration and performance). %Subjects can rate each parameter on a 7-point Likert scale \cite{joshi2015likert} where 1 (very low) and 7 (very high). For evaluation, there are number of variants are available depending upon the task.  The details about the dimensions are listed below.
%\begin{itemize}
%\item Mental Demand: how much thinking, deciding, or calculating was required to perform the task
%\item Physical Demand: the amount and intensity of physical activity required to complete the task
%\item Temporal Demand:  the amount of time pressure involved in completing the task.
%\item Efforts: how hard does the participant have to work to maintain their level of performance?
%\item Performance: the level of success in completing the task.
%\item Frustration Level: how insecure, discouraged, secure or content the participant felt during the task.
%\end{itemize}
This implicit evaluation enables us to examine 
%understand 
the cognitive load and cognitive engagement of the user while using a system.

\end{enumerate}

\begin{table}[]
    \centering
    \begin{tabular}{|c|c|c|c|}
    \hline 
& Negative	&  Scale & Positive \\ \hline 
& obstructive & 1 2 3 4 5 6 7 & supportive \\ 
Pragmatic quality  & complicated & 1 2 3 4 5 6 7 & easy \\
& inefficient & 1 2 3 4 5 6 7 & efficient \\
& confusing & 1 2 3 4 5 6 7 & clear \\  \hline
 & boring & 1 2 3 4 5 6 7 & exciting \\
Hedonic quality & not interesting & 1 2 3 4 5 6 7 & interesting \\
& conventional & 1 2 3 4 5 6 7 & inventive \\
& usual	& 1 2 3 4 5 6 7 & leading edge \\ \hline
    \end{tabular}
    \caption{Scales pragmatic quality and hedonic quality}
    \label{tab:User Experience}
    %\vspace{-2ex}
\end{table}

\subsubsection{Contentment Segment:}

This segment focuses on information need satisfaction during the search process. It includes a questionnaire based on interaction while searching, learning during searching and knowledge gain arising from the search activity:

\begin{enumerate}
 % Learning is defined as change in the pre-extinguish knowledge structure. 
%Vakkari \cite{vakkari2016searching} explained in detail how learning occurs in the search process. It also described the change in the knowledge structure related to accessing and interacting with the search engine or knowledge source. %According to study \cite{vakkari2016searching}, search stage and knowledge structure are interrelated. Search stage classified into three dimensions: search formulation, source selection and interacting with sources. Similarly change in knowledge structure is divided into three phases restructuring, tuning and assimilation.  

\item Search as Learning: As discussed in Section \ref{Essential Factors for Conversational}, it is important to observe whether a CS
%conversational search 
system 
%is
supports the user effectively in their engagement 
%interacting 
with the search system, and enables the user's 
%to change its 
knowledge gain arising from the search process.
%structure (gain). 
To better understand this process,
%For better understanding, 
we decided to separately measure 
%both 
the factors of 
%both 
%such as 
user interaction and modification of their mental knowledge structures.
%Based on user studies and reviewing multiple research literature, 
We developed
%drafted 
a questionnaire \cite{Abhi}, as shown in Table \ref{FLowchart}, to observe user interaction behaviour, 
%(search stage) 
inspired by Vakkeri's model of search as learning\cite{vakkari2016searching, vakkari2003changes}.
%This questionnaire helps us to study the interaction and search activity of users while changing the knowledge structure as shown in Table \ref{FLowchart}. Regarding other dimensions, change in the knowledge structure, we suggested another metric in the next section that enable to investigate the change in knowledge structure while searching.      
%Based on the search as learning Vakkeri model \cite{vakkari2016searching}, the user search experience can be evaluated on 15 parameters including the relevance of the search result \cite{Abhi}, the quality of the text presented by the interface, and understanding of the topic in both the search settings via pre-search and post-search questionnaires.

\item Knowledge Gain: %Search allows one to change its knowledge structure but its very difficult to empirically measure it. % Generally, Conversational search are exploratory in nature and the current metric has been introduced accordingly.
To measure the knowledge gain, the user 
%need to 
is required to write a pre-search summary and a post-search summary relating to 
%about 
the search topic. This summary is
%will be
manually evaluated by an independent assessor on three sub dimensions: %as described %mentioned 
%in 
%the study by Wilson et. al. 
 %The criteria are:
 Quality of Facts (Dqual), Interpretation (Dintrp) and Critiques (Dcrit), as shown in Table \ref{Summary Comparison Metric}\cite{wilson2013comparison}.
This difference between these summaries indicates knowledge gain from the search process.
%The summary was scored against these three factors by two independent analysts with the Kappa coefficient (Approx .85) \cite{Kappa}
%\ref{tab:Summary Comparison Metric}
 %Knowledge expansion was investigated using a comparison of pre-search and post search summaries based on a number of parameters, as shown in Table \ref{Summary Comparison Metric}, while using both the systems, we divide the hypothesis into two sub-parts as follows: 

\begin{table}[t]
    \centering
    \begin{tabular}{l|l}
    \hline
    Parameter & Definition \\
    \hline
   Dqual & Comparison of the quality of facts in the summary in range 0-3, where 0 \\ & represents irrelevant facts and 3 specific details with relevant facts. \\
   \hline 
   Dintrp & Measures the association of facts in a summary in the range 0-2, where 0\\ & represents no association of the facts and  2 that all facts in a summary are\\ & associated with each other in a meaningful way. \\
      \hline 
   Dcrit & Examines the quality of critiques of topic written by the author in\\ & range the 0-1, where 0 indicates that
   %represents 
   facts are listed without analysis\\ & and 1 where both advantages and disadvantages of the facts are given. \\
   \hline
    \end{tabular}
    \caption{Summary Comparison Metric \cite{wilson2013comparison}}
    \label{Summary Comparison Metric}
%    \vspace{-2ex}
\end{table}

%\begin{enumerate}
 %   \item Comparison of
    %the 
  %  pre-search and post search summaries: This is to verify the knowledge expansion after each task independent of the search interface used by the user.
   % \item Comparison of the mean difference between
    %of 
%    pre-search and post-search summary for 
    %of 
 %   each interface: This is to verify which interface supported users better in gaining 
    %the 
  %  knowledge. 
%\end{enumerate}

%The user gains knowledge after the post search in both the search interface.  %The objective of this hypothesis testing is to find that subjects were gaining knowledge while doing the search independently from  the type of interface. 
%The knowledge gain was examined by analyzing the pre-search 
%summary 
%and post search summaries. We 
%have 
%asked subjects to write a short summary of the topic before the search and after the search. The summary was
%has been 
%analyzed based on three 
%important 
%criteria as described in
%mentioned by the author 
%\cite{wilson2013comparison}. The criteria are: Quality of Facts (Dqual) , Intrepterations (DInterpreataion) and Critiques (DCritique), as shown in Table \ref{Summary Comparison Metric} 
%\ref{tab:Summary Comparison Metric}. 
%The summary was
%has been coded on 
%scored against
%all 
%these three factors by two independent analysts 
%coders 
%with the Kappa coefficient (Approx .85) \cite{Kappa}. We conducted 
%the 
%hypothesis T dependent testing on tasks completed using both the 
%search task 
%conventional
%Traditional 
%search interface
%task 
%and the MCSI
%Chat
%Search Task
%).

\end{enumerate}

\begin{figure}[h]
    \centering
    \includegraphics[scale=0.55]{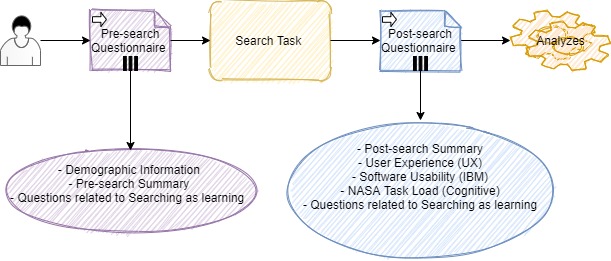}
    \caption{Evaluation Process}
    \label{fig:Evaluation Process}
     \vspace{-2ex}
\end{figure}

\subsection{Developing the Evaluation Process}

The overall evaluation process is 
%can be developed as 
shown in the Figure \ref{fig:Evaluation Process}. 
%The 
The user
%Users 
completes 
%have to fill the 
a pre-search questionnaire and then %complete search task and then filled 
%and the 
a post-search questionnaire. The questionnaire is
%has been designed 
based on the metric discussed in Section \ref{coneptual}, and as shown in Figure \ref{fig:Metric}. To maintain
%keep the 
uniformity, subjects 
%can 
rate each parameter on a 7-point Likert scale \cite{joshi2015likert}, where the scale ranges from 1 (very low) to
%and 
7 (very high) on each questionnaire. The evaluation is conducted from
%can be conducted in 
two perspectives: a) comparison of a conversational interface with a conventional search system, b) evaluating only a conversational interface based on a provided benchmark. Details are given 
%explained 
below:

\begin{itemize}

\item Comparison of conversational interfaces with conventional search systems: This evaluation method enables 
%one to compare 
comparison of 
%the 
conventional and conversational search interfaces based on 5 dimension metrics, as discussed in Section \ref{coneptual}. The user 
%need to 
completes two search tasks, 
%based 
one 
%each 
using 
%on 
each 
%different 
of the search interfaces
%setting 
(a conventional and a conversational interface). For each task, the user completes a 
%need to fill 
pre-search questionnaire and a post search questionnaire. 
%And should observe the difference in each dimension in both the setting (conversational and convectional search system).
%The inference forms the results would assist in challenges and opportunities in the conventional search system. 
This analysis is intended
%inference might help 
to provide better insights into 
%vision for 
the operation of a
%the 
CS 
%conversational search 
system and contrasting user opinions of each type of interface.
%. The investigation would also lead to understand the user opinion in both types of the search system. 

\item Evaluating 
%only the 
a conversational interface based on standard numerical benchmarks: Most of the metrics introduced 
%discussed 
in the framework have 
%their 
a standard numerical benchmark \cite{Abhi} \cite{wilson2013comparison} \cite{hart1988development} \cite{lewis1995ibm}. Pre-search and post-search questionnaire scores for
%of 
each dimension of evaluation can be compared using 
%with 
their standard benchmarks. This %would 
not only provides an
%rough 
%estimation 
estimated evaluation of its effectiveness, but also 
%it would 
provides an opportunity to explore the conversational interface with the standard system benchmarks. Furthermore, this 
%would 
allows us to understand 
%the 
user expectation in general in all dimensions. This provides 
%the
empirical measurability of the CS
%a conversational search 
interface in the light of the benchmark. Moreover, this can
%could
help us to understand how far from or close to, the current CS
%conversational search 
interface is to 
%with respect to 
the user's expectations.   
\end{itemize}

As per above the 
%mentioned 
prospective, it is very important to analyse the data critically
%. These critical analyses could be tested statistically 
including 
%to confirm 
use of statistical significance tests.
%results.
If the results are significant, this
%that it may 
can be used to develop a separate benchmark for the CS
%conversational search 
interface to
%which would 
assist other researchers in comparing
%to compare 
their studies on CS
%conversational 
%search 
interfaces. 

\subsection{Implementation and Analysis of the Framework}

As mentioned earlier, the user is required to complete 
%must fill the  
pre-search and post search questionnaires. 
%Based on the implicit feedback, the conversational interface will be evaluated on the framework. The framework would indicate the conversational interface performance into all five dimensions. 
%The feedback collection process has been breakdown into two parts i.e. collecting data via pre-search questionnaire and post-search questionnaire. 
We have developed 
%The 
%We have drafted the prospective 
prototype 
%these 
questionnaires by 
%after 
combining 
%all 
the dimensions mentioned earlier in Section \ref{coneptual}\footnote{The 
%respective 
questionnaires can be found at
%reached by the link 
\url{https://forms.gle/MaooazzEfQJ4sTpPA}}. The details of
%about 
our
%the 
pre-search 
%questionnaire 
and post-search questionnaires are
%is 
described below.  

\begin{enumerate}

\item \textbf{Pre-search Questionnaire:}
This 
%section 
focuses only on contentment, and contains 
%carries 
questions on demographic details of searcher, background knowledge of the searcher about the search topic, interest in the search topic, searcher experience of using conversational systems, etc.  

\item \textbf{Post-search Questionnaire:}
This 
%section 
focuses on contentment and exploration, and contains 
%carries 
questions on knowledge gain after search, based on interactions (e.g., How many documents reviewed by user?), software usability, UX, cognitive load, etc. 

\end {enumerate}

\vspace{-2ex}

\begin{figure}
    \centering
    \includegraphics[width=\textwidth, height=8cm ]{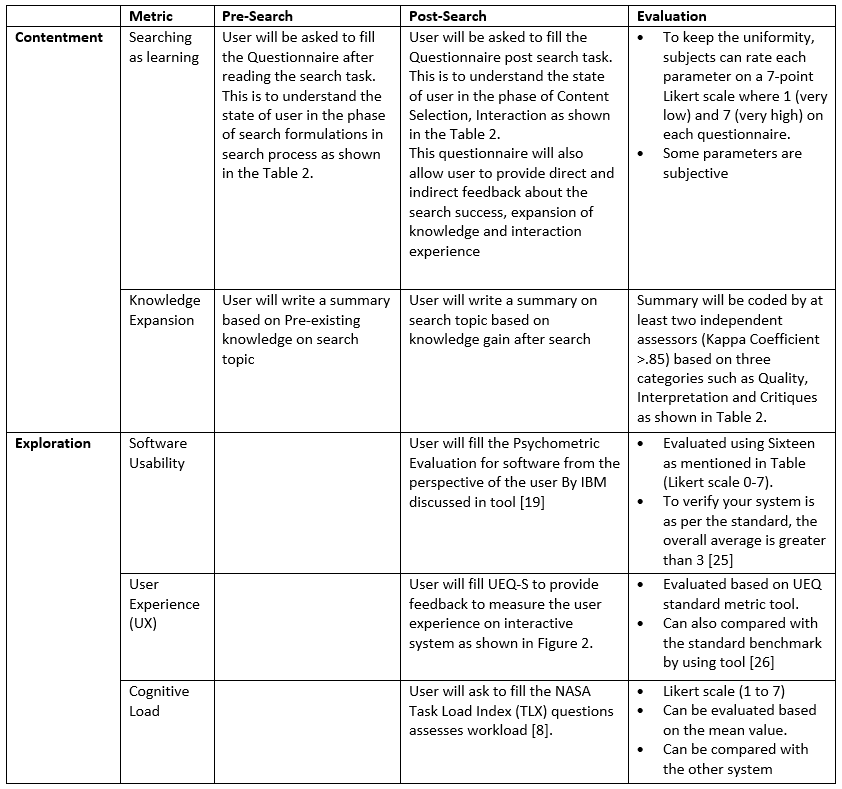}
    \caption{Implicit Evaluation for Conversational Search Interface Metric.}
    \label{fig:Metric}
\end{figure}

Each question is
%would be 
evaluated based on a Likart score [0,7],
%(0,7), 
except for the knowledge gain metric. As described, the framework is classified into two sections: exploration and contentment, as shown in Figure \ref{fig:Metric}. Details are as follows:

\begin{enumerate}
    \item Exploration : The questionnaire to investigate exploration is aligned to the user based on 
    %after 
    their search experience. As such,
    %So, 
    the conversational interface is evaluated based on the post-search questionnaire. The mean score of each question is calculated based on the number of users. 
    %The 
    Analysis is conducted using both Quantitative Analysis and Qualitative Analysis.
    \label{Exploration}
    \begin{enumerate}
        \item Quantitative Analysis: This is based on the mean score of the participants in the study, statistical testing is carried out based on the population and nature of the experiment. When comparing a conventional system and a conversational system, we are able to perform dependent significant testing, since the population undertaking the experiment in both settings is the same. And, if we are comparing the mean score of the conversational interface with a standard benchmark, 
        %then 
        we can conduct independent significance testing. This statistical testing enables us to understand how 
        %our 
        systems differ.
        %is different and better. 
        Additionally, each dimension discussed above in Section \ref{coneptual} has a 
        %their 
        standard tool for analysis. 
        % ??? Those tool could also be explored.  
        \item Qualitative Analysis: The different dimensions 
        %zones 
        are annotated based on comparison of the mean values for the study participants.
        %comparison.  
        A mean value between 2 and 4 represents a neutral evaluation of the corresponding scale (yellow dimension),
        %zone), 
        a mean $>$ 4 represents a positive evaluation (green dimension)
        %zone)
        and  mean $<$ 2 represents a negative evaluation (blue dimension).
        %zone). 
        After comparing the mean, each question is annotated based on the dimensions.
        %zone. 
        The dimensions
        %zones 
        are annotated by
        %based on 
        two independent analysts with the Kappa coefficient (Approx .85), then the dimension
        %zone 
        is counted for each section such as software usability, user experience \cite{sauro2016quantifying,UEQ}, cognitive load. As per the dimension,
        %zone, 
        the section of the interface that needs to be improved can be identified. For example, if software usability gets more red dimensions,
        %zones, 
        then the interface needs to be improved with respect to software usability. %In each question in the questionnaire, the grading scale should carry the same standard where 0 means negative and 7 means positive.   
    \end{enumerate}
    
    %Based on the mean comparison, the different zones would be annotated. Mean between 2 and 4 represent a neural evaluation of the corresponding scale (yellow zone), Mean $>$ 4 represent a positive evaluation (green zone) and Mean $<$ 2 represent a negative evaluation (green zone).  After comparing the mean, each question would be annotated based on zone. The zones need to be annotated based on two independent analysts with the Kappa coefficient (Approx .85). Then zone would be counted for each section such as software usability, user experience \cite{sauro2016quantifying,UEQ}, cognitive load. As per the zone, it could be identified in which section the interface needs to improve. For example, if software usability gets more red zones, then the interface needs to be improved in software usability. In each question in the questionnaire, the grading scale should carry the same standard where 0 means negative and 7 means positive.     

\item Contentment: The questionnaire to investigate contentment is aligned to the user's pre-search knowledge and post-search knowledge. As discussed earlier, contentment evaluation is 
%has been 
designed to investigate user learning while searching, and their knowledge expansion arising from
%after 
the search process. The analysis can again be conducted using both Quantitative Analysis and Qualitative Analysis.

\begin{enumerate}
    \item Quantitative Analysis: %If we are comparing the convectional system and conversational system then we can do dependent significant testing, as population undertake the experiment in both setting would be same. 
    Based on the mean score for the study participants of searching as learning and knowledge gain (the difference
between pre-search and post-search summaries of each setting (conventional system and conversational system)) parameters, statistical testing can be applied. %And if we are comparing the mean score of conversational interface with standard benchmark then we could conduct the independent significant testing.
    \item Qualitative Analysis: Search as learning questions, as shown in Table \ref{FLowchart}, are annotated, evaluated and analyzed based on different dimensions as discussed in the previous section on Qualitative Analysis in Exploration (corresponding scale (yellow, green and red dimensions)). Pre-search and post-search summaries can 
    %should 
    be compared based on the parameters discussed in Table \ref{Summary Comparison Metric}. %This metric is not based on a Likert scale. 
    The summary is scored against all these factors by two independent analysts with the Kappa coefficient (Approx 0.85) \cite{Kappa}. For each parameter, the difference between pre-search and post-search summaries is calculated. If the difference of Dqual$>$ 1.5, Dintrp $>$ 1 and Dcrit $>0$. it is assumed that the user has increased their knowledge by more than 50\%.
    
\end{enumerate}

 %Then zone would be counted for each section. As per the zone, it could be identified in which section the interface needs to improve. This investigation would validate that the user has the learning experience while the search process. And it’s very difficult to calculate the quality of the learning. 
%The user is asked to write the pre-search summary on the search topic based on his or her background knowledge. User is also asked to write the post search summary based on its learning. 

\end{enumerate}

%\section{Conclusions and Future work}
\section{Concluding Remarks}

The concept of conversational search (CS) remains an ongoing topic of research. A crucial part of this work is 
%As part of these considerations, 
evaluation of CS.
%is a critical consideration of how to advance research in CS. 
Studies 
%based on 
of CS 
%have 
to date have mainly been based on user experience.
%software usability and cognitive load. 
This overlooks interaction with the system and changes in the user's knowledge structure. %In the paper, we have studied the evaluation methods of interactive information retrieval, conversational system, conversational search. 
In this paper, we examine the factors that can impact on the effectiveness of a CS interface. Following our investigation, we propose an evaluation framework for CS that incorporates
%satisfies 
the evaluation methods from 
%in 
interactive IR and conversational systems. We believe 
%that 
our proposed evaluation framework for CS to be practical and applicable in real life scenarios, and will provide greater insights for understanding and advancing CS processes than is possible using the evaluation methods used in existing work on CS. 
%Our framework is a generalized combination of many evaluation methods already used for the evaluation of conversational systems and interactive IR systems. 
We are currently working on the 
%plan to 
validation of our proposed framework within our ongoing study of CS. 
%The conversational search is a new domain of information retrieval. 
%It may be the case that the conversational search agent would be evaluated based on some variables presented in the study and may also include some new variable of evaluations. We believe that the whole research community would work together to define conversational information retrieval and its evaluations methods . 

% \usepackage{multirow}
% \usepackage{multirow}

\section*{Acknowledgement}

This work was supported by Science Foundation Ireland as part of the ADAPT Centre (Grant 13\//RC\//2106) at Dublin City University.

\bibliographystyle{splncs04}
\bibliography{ref}
\end{document}